\documentclass{acm_proc_article-sp}
\usepackage{helvet}
\usepackage{courier}

\usepackage[hyphens]{url}      
\usepackage{balance}  
\usepackage{graphicx} 
\usepackage{times}    
\usepackage{multirow}
\usepackage{float}
\usepackage{caption}
\DeclareCaptionType{copyrightbox}
\usepackage{subcaption}

\usepackage{algpseudocode}
\usepackage{algorithm}

\begin{document}



\title{Co-Following on Twitter}

\numberofauthors{2} 
\author{
%
%
\alignauthor Venkata Rama Kiran Garimella\\
\affaddr{Qatar Computing Research Institute}\\
\affaddr{Doha, Qatar}\\
\email{gvrkirann@gmail.com}
			%
\alignauthor Ingmar Weber\\
\affaddr{Qatar Computing Research Institute}\\
\affaddr{Doha, Qatar}\\
\email{iweber@qf.org.qa}
}

\maketitle
\begin{abstract}
We present an in-depth study of co-following on Twitter based on the observation that two Twitter users whose followers have similar friends are also similar, even though they might not share any direct links or a single mutual follower. We show how this observation contributes to (i) a better understanding of language-agnostic user classification on Twitter, (ii) eliciting opportunities for Computational Social Science, and (iii) improving online marketing by identifying cross-selling opportunities.

We start with a machine learning problem of predicting a user's preference among two alternative choices of Twitter friends. We show that
 co-following information provides strong signals for diverse classification tasks and that these signals persist even when (i) the most discriminative features are removed and (ii) only relatively ``sparse'' users with fewer than 152 but more than 43 Twitter friends are considered.

Going beyond mere classification performance optimization, we present applications of our methodology to Computational Social Science. Here we confirm stereotypes such as that the country singer Kenny Chesney (@kennychesney) is more popular among @GOP followers, whereas Lady Gaga (@ladygaga) enjoys more support from @TheDemocrats followers.

In the domain of marketing we give evidence that celebrity endorsement is reflected in co-following and we demonstrate how our methodology can be used to reveal the audience similarities between Apple and Puma and, less obviously, between Nike and Coca-Cola. Concerning a user's popularity we find a statistically significant connection between having a more ``average'' followership and having more followers than direct rivals. Interestingly, a \emph{larger} audience also seems to be linked to a \emph{less diverse} audience in terms of their co-following.


\end{abstract}




\section{Introduction}\label{sec:intro}




How much does following a particular set of people reveal about your interests? Does the fact that you follow @Starbucks make it more likely that you follow @TheDemocrats as well? And can Twitter users be grouped in a meaningful way by looking at whether their followers have similar friends\footnote{We use the term ``friend'' as Twitter terminology referring to another Twitter user that a user follows.}?

Such questions are relevant to at least three lines of research. First, there is lots of work on user classification on Twitter, e.g., \cite{cohenruths13icwsm,conoveretal11socialcom,barbera13polnet}. Such classifiers often rely on language-specific tools such as stemming or dictionaries with special terms. Our work shows that such information might not be required and a language-agnostic method using a user's friends as features achieves AUC-ROC of .80-.85 for a wide range of binary classification tasks. 
Second, online social networks are becoming a more and more important data source for Computational Social Science~\cite{lazer2009life,cioffi2010computational}. We contribute to this area by showing how things such as ``lifestyle politics'' can be studied by using co-following information.
Lastly, Twitter with its user base of several hundreds of millions is an important advertising and marketing platform. We show how followership-based similarity methods can be used to identify accounts with a similar audience in terms of interests which could create cross-selling opportunities.

\begin{figure}
\centering
\begin{subfigure}{0.23\textwidth}
\includegraphics[width=\textwidth, clip=true, trim=0 400 0 0]{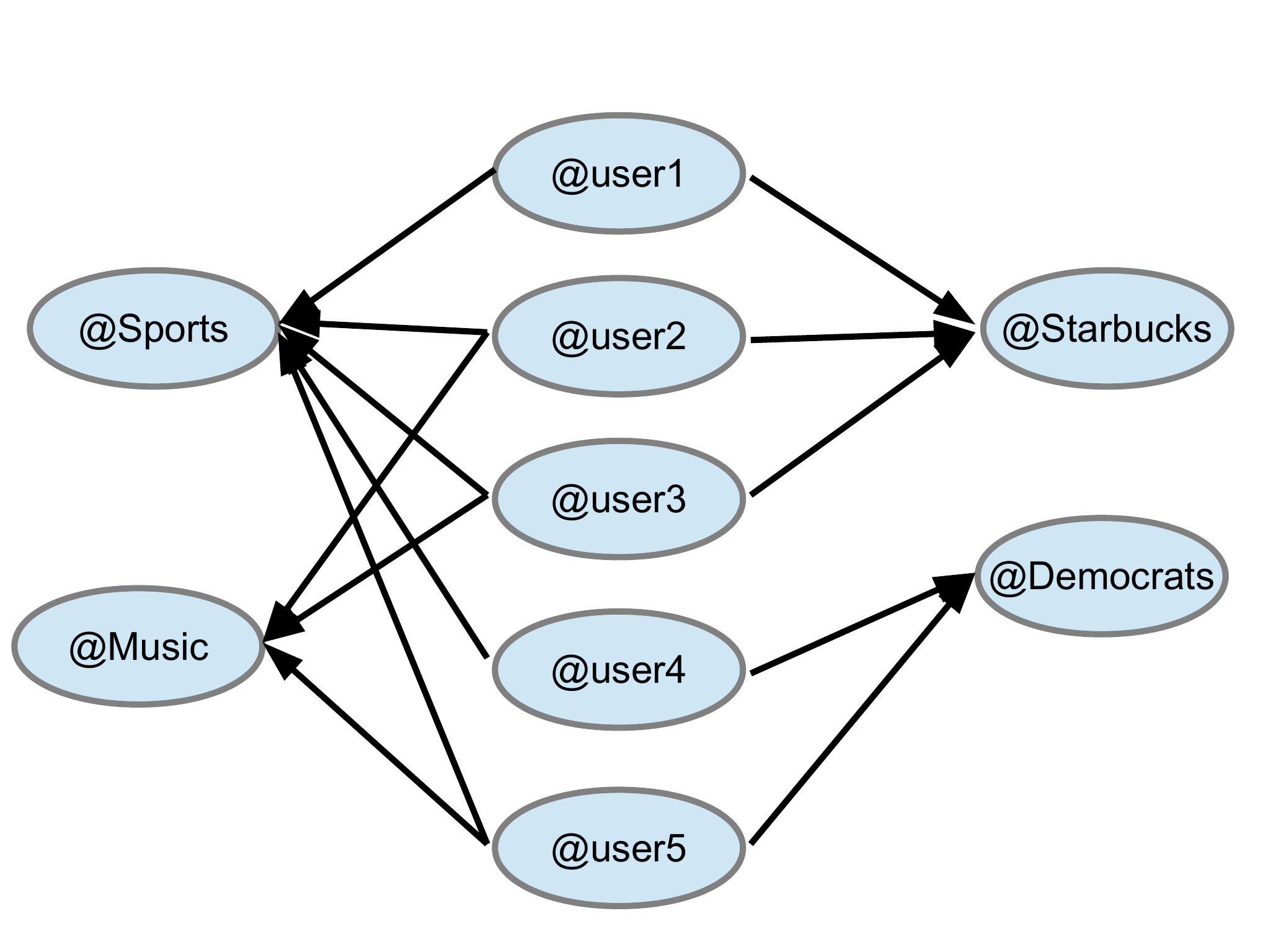}
\caption{}
\label{fig:cofollowing}
\end{subfigure}
\begin{subfigure}{0.23\textwidth}
\includegraphics[width=\textwidth, clip=true, trim=0 400 80 0]{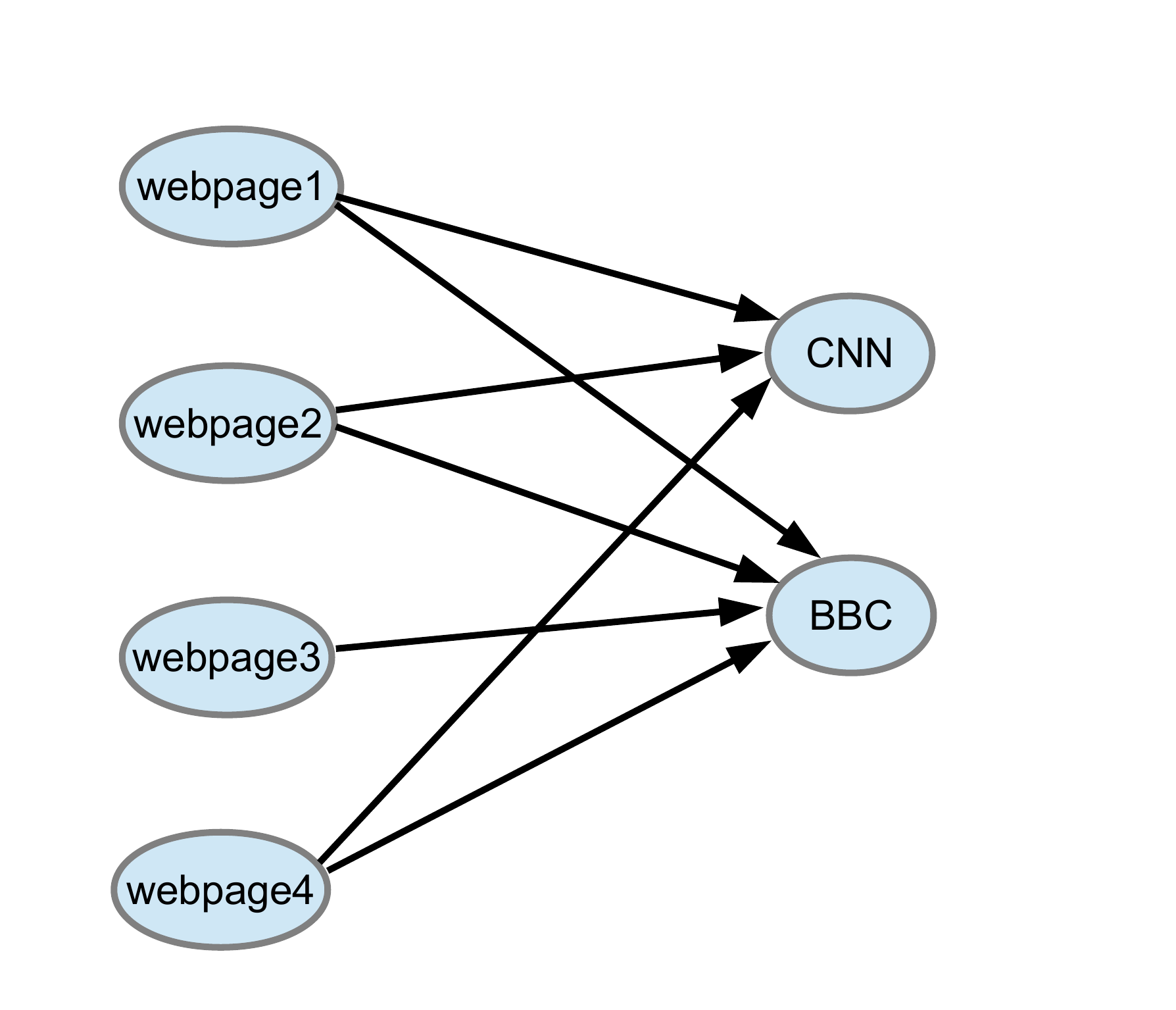}
\caption{}
\label{fig:hubs}
\end{subfigure}
\caption{(\ref{fig:cofollowing}) - Starting from @Starbucks and @Democrats, our co-following analysis uses both the 1st-hop backward edges to the users and the 2nd-hop forward edges to other co-followed accounts. (\ref{fig:hubs}) - Traditional co-linkage clustering of webpages and links would focus on direct 1st-hop backward edges and use shared inlinks to detect similar webpages, in this example CNN and BBC.}
\label{fig:hubs_cofollowing}
\end{figure}

As an example of our aproach, consider Figure~\ref{fig:hubs_cofollowing}. Here Figure~\ref{fig:cofollowing} shows the idea that we propose. As an example, consider the twitter accounts of Starbucks (@Starbucks) and the Democrats (@Democrats). Directed edges from a users (user1-user5 in the middle) indicate following behavior of these users. In this example, we see that the two accounts of interest do not share a single common follower. However, many of their followers are ``co-following'' other, seemingly unrelated accounts. This can be used to deduce the similarity or rather closeness of the two accounts. At first sight, this idea is similar to using common links (co-citations) for clustering web pages~\cite{small1973cocitation,wijaya2006clustering}. E.g., in Figure~\ref{fig:hubs}, the webpages on the right, CNN and BBC, do not have any links between each other, but could be grouped together using similar links from webpage1 - webpage4. However, typical co-citation or co-linkage approaches would focus on the ``1-hop backward'' links only and then looking at overlaps. In our analysis, we make crucial use of the added ``forward'' links. In a sense, we are using 2nd order co-citation or co-following rather than ordinary 1st order co-citation.

Most of our analysis is centered around 18 rivalries such as @GOP vs.\ @TheDemocrats or @McDonalds vs.\ @BurgerKing. In many cases the two alternatives are arguably interchangeable and one might not expect a big difference between the interests of the followers of, say, @Hertz and @Avis. For each of these seed pairs we obtained up to  2,000 random followers. Their friends are used to construct feature vectors and we perform an in-depth analysis of the co-following behavior. We also construct the same kind of vectors for a set of (i) French and German political parties, and (ii) popular musicians. In all cases the basic hypothesis is that users following similar users are similar and that this propagates to their friends.
Concretely, we investigate if users whose followers have similar friends are also similar, even if they might not share any direct followers.

Our findings include the following.


\begin{itemize}
\item Using \emph{solely} language-agnostic co-following information provides strong signals concerning a user's preference even among arguably interchangeable choices such as @Hertz or @Avis.  

\item Such classification is robust with respect to the removal of the most strongly and often obviously related co-following features.  

\item Aggregated signals from the general crowd work better for distinguishing binary preferences than relying on the most similar users in a k-NN fashion.
For discerning the ``center'' dual-follower users from the rest the opposite holds. 

\item Following 50-150 users suffices for a strong signal with little additional gains beyond that. 

\item A feature analysis confirms stereotypes such that @ladygaga is more popular among @TheDemocrats followers, but also reveals less expected patterns such as that @SnoopDogg followers tend to prefer @Pepsi over @CocaCola. 

\item There is evidence that celebrity endorsement works as following a celebrity increases the probability of following the related product. 

\item Users following \emph{both} alternatives of two rivals are sometimes ``in the middle'' and sometimes a ``class by themselves'', thus making it harder to detect such users than either of the two poles. 

\item Groups of related Twitter accounts, such as musicians or political parties, can be mapped in a simple manner by looking at their followers' friends. Such a mapping reveals interesting facts, for example, that Apple and Puma target a similar, metropolitan audience. 

\item There is statistically significant evidence that Twitter users whose followership is closer to a global ``average'' have more followers than their rivals.  

\item There is a positive effect on the total follower count by having more intrinsically similar and coherent followers.

\end{itemize}

To the best of our knowledge, this is the first such study of co-following on Twitter. We hope that both our analysis and our tools will be of interest to researchers working on user classification, Computational Social Science, or on Social Media marketing.

The rest of this paper is organized as follows. First we review related work from domains such as Twitter user classification or follower recommendation. Our data set and pre-processing steps are described in the following section, Section~\ref{sec:data}. 
In Section~\ref{sec:binary} we investigate the potential of co-following information for discerning preferences among similar but rivaling accounts.
In Section~\ref{sec:binary:middle} we show how co-following information can be used to group accounts with similar followership in a meaningful manner. 
In Section~\ref{sec:mapping} we look at how co-following based similarity can be used to visually discriminate between various accounts from domains as diverse as politics and music.
Finally, we look at whether it pays off to have mainstream followers in Section~\ref{sec:mainstream} before concluding in Section~\ref{sec:conclusions}.

%
\section{Related Work}\label{sec:related}

A key assumption to group users based on the similarity of their followers' friends is that following is an expression of topical interests or demographic similarity, rather than personal contacts. 
Kwak et al.\ address the question whether Twitter is more of a social network or a news media \cite{kwaketal10www}. Looking at trending topics they find ``that the majority (over 85\%) of topics are headline news or persistent news in nature''. They also observe a low degree of reciprocity among links, which is atypical for normal social networks. Note that for our applications it is \emph{good} that links on Twitter are not dominated by real social ties as this indicates that they are more related to interests. 
%
%
An early study of why people use Twitter is presented in \cite{javaetal07webkdd}. The authors find ``that people use microblogging to talk about their daily activities and to seek or share information''. Given the findings of Kwak et al., the information sharing fraction has likely increased since which, again, benefits our approach. 
A similar conclusion ``that Twitter users have a very small number of friends compared to the number of followers and followees they declare'' is reached in \cite{huberman09fm}, further indicating that the majority of follow links is likely related to topical interests, not real-life personal connections.
This observation is also confirmed in \cite{weietal12websci} where Wei et al.\ find that mutual following is a signal for real life friendship. 
%

Related to our approach of using co-following to measure user similarity is the work in \cite{anetal11icwsm} where An et al.\ use the common audience of two Twitter accounts as a measure of closeness. 
This differs fundamentally from our methodology as we use \emph{second order} co-following. Concretely, two accounts that do not share a single follower are considered similar by us if their followers share many friends. 
We believe that such an approach is preferable to break out of the ``homophily ghetto'' where users follow what their friends follow. It also allows for a more far-reaching notion of similarity that could be used to, say, align 
political parties in different countries on a common spectrum, even if no user follows parties in two distinct countries. The benefits of using higher-order co-occurrences rather than only direct co-occurrences have been shown for document retrieval via
dimensionality reduction \cite{bastmajumdar05sigir}.
Outside the scope of Twitter, authors have also studied first order co-citation and co-linking patterns. 
Small introduces the idea of looking at co-cited academic papers in \cite{small73jasis} with the aim of ``provid[ing] a new way to study the specialty structure of science''. 
Culnan looks at author similarity in the realm of Management Information Systems by looking at which authors get co-cited \cite{culnan1987mapping}. 
Common, direct inlinks are used as one of the signals in \cite{wangkitsuregawa01lncs} to form the basis of webpage clustering.
%


Several papers have looked at various Twitter user classification tasks, typically for (i) political orientation in the US, (ii) gender, and (iii) age \cite{pennacchiottipopescu11kdd,raoetal10smuc,alzamaletal12icwsm,cohenruths13icwsm,conoveretal11socialcom,barbera13polnet}. 
This line of work usually involves a broad set of features, including textual content, network and activity based features, as well as a variety of classification approaches that make use of label-propagation across social links. 
%
%
%
%
%
Our approach in Section~\ref{sec:binary} 
differs in a number of ways. First, the binary classification task of ``does the user follow A or B'' is different. Second, we do not use any content-based features. Third, we do not use any retweet signals as we are not interested in the sparse network of interests that users strongly engage with, but rather the larger network of ``weak interests''. Fourth, we do not make use of label-propagation across social links as we are not interested in methods that work for (and re-inforce) ``information bubbles'' but we are looking for approaches that can be transferred to completely new domains where users do not yet have any direct social ties. 
Finally, the actual classification performance is of less interest to us than an understanding of how much information is contained in co-following and how this could be used for different applications.


Given its global popularity and the relative ease of data acquisition through open APIs\footnote{\url{https://dev.twitter.com/}}, Twitter has been used before for studies in Computational Social Science and some examples are discussed in the following. 
Quercia et al.\ use a crawl of 228K Twitter profiles to test ``whether established sociological theories of real-life networks hold in Twitter''  \cite{querciaetal12icwsm}. They find that, ``much like individuals in real-life communities,  
social brokers [...] are opinion leaders who tweet about diverse topics, have geographically wide networks, and express not only positive but also negative emotions''.  
%
%
Garcia-Gavilanes et al.\ link cultural variables such as Pace of Life, Individualism and power to tweeting behavior from several countries \cite{garciaetal13icwsm}. They find strong correlations between these variables and online behavior. 
Though our current paper only partly falls under the  umbrella of Computational Social Science, we hope that this work still illustrates the potential of using co-following information for such studies.

Our mapping and visualization of similar accounts in Section~\ref{sec:mapping} is conceptually similar to community detection which also identifies groups of related accounts.
Classical community detection looks at the global network and tries to find areas with unusually high triadic closure \cite{newman06pnas} or otherwise unusual linkage patterns.
A survey of the area and experimental comparisons can be found in \cite{fortunato10pr} and \cite{lancichinettifortunato09physreve} respectively. Related are notions of graph centrality that also require the whole graph \cite{freeman77sociometry}. 
The approach in \cite{panetal10pysica} is arguably the most similar to ours as it combines local structural information with node similarity in an iterative manner. The similarity used, however, requires direct co-links and even direct links. 
Our approach differs in a number of ways. First, we are not interesting in clustering/grouping \emph{all} users but only understand the relative positions of main ones. Second, we do not require a global view of the network but, in return, we cannot argue about things such as node centrality or PageRank. Third, we do not want to find communities induced by friends-of-friends type links. Rather we strive for a similarity-only based approach that can easily be transferred to domains without any friends-of-friend links.

Determining which of two alternatives a Twitter user is more likely to follow is related to friend recommendation or link prediction as, in a sense, we are suggesting which of the two links should be formed. 
Intuitively, transitivity and mutuality of links are important signals for link prediction \cite{golderyardi10socialcom}  but, as discussed previously, we do not want to use such ``three people you follow also follow X'' information as it leads to a different type of application, closer to community detection. User similarity based on user attributes has also been used as a feature for link prediction \cite{yinetal10asonam,yinetal11cikm,huttoetal13chi}. But this work still partly relies on mutuality and transitivity, which is equivalent to re-enforcing partisan camps without noting any existing similarities in terms of shared interests. For example, such approaches would most likely fail to pick up the similarity between @Puma and @TheAppleInc that we observe in Section~\ref{sec:mapping:all_rivals}.

Related to our analysis in Section~\ref{sec:mapping} on mapping and visualization of similar users is previous work on segmenting online user populations as our methods can also be applied for audience analysis on Twitter. 
For web search a clustering based market segmentation is presented in \cite{weberjaimes11wsdm}. The authors include demographic variables in their analysis which could also be obtained for Twitter through the help of machine learned classifiers. 
For web browsing a similar study, also including demographic variables, is presented in \cite{goeletal12icwsm}. 
In our work we do not explicitly segment millions of Twitter users but, rather, use regular Twitter users to segment a small number of important accounts.

As mentioned in Section~\ref{sec:intro}, co-citations and common links have been used as a measure of similarity of scientific articles and webpages for a long time~\cite{small1973cocitation,wijaya2006clustering}. The idea there is that if two webpages have similar inlinks (think ``endorsing followers''), they are similar. Our idea is similar to these in the sense that we are using the notion of ``backward'' links (pages linking to = authors citing = users following) to look for similarity, but it differs as (i) we not only make use of the 1st-hop backward links, but also use 2nd-hop forward links (other accounts followed by a user) to measure similarity. Finally, our application scenario is different as we are not interested in finding, say, topically similar web pages but, rather, trying to find hidden closeness between a priori unrelated accounts on Twitter.

\section{Data}\label{sec:data}
Our data set is constructed around a set of \emph{Twitter seed accounts}. These accounts correspond to ``rivalries'' between two entities such as @CocaCola vs.\ @Pepsi or @Samsung vs.\ @TheAppleInc.
The full list of 18 account pairs can be found in Table~\ref{tab:performance}. 
The list of these rivalries was obtained from\footnote{\url{http://money.cnn.com/gallery/news/companies/2013/03/21/greatest-business-rivalries.fortune/index.html}}. We only considered those rivalries that are relevant and had a twitter account.
To study the ``middle'' users in Section~\ref{sec:binary:middle}, we randomly sampled a set of 2,000 users who follow \emph{both} rivaling accounts. 
Later, we also look at \emph{groups} of seed accounts, namely, Twitter accounts for (i) German and French political parties, (Section~\ref{sec:mapping:france_germany}),
(ii) popular musicians, (Section~\ref{sec:mapping:musicians}),
and (iii) all the 18 rivalries combined, (Section~\ref{sec:mapping:all_rivals}). 
In all cases, we first obtained a list of all the accounts' followers. From this list we then sampled uniformly at random a set of 2,000 followers. For the cases of rivalries, we also imposed the constraints that the followers were located in the United States. This was done to avoid picking up differences in international market penetrations, rather than within-US cultural differences. The followers for the political parties from Germany and France
 were also limited to Germany and France respectively. 
To implement the within-country restriction, we first sampled a set of 10,000 followers for each seed user uniformly at random (or using all when the account had fewer followers), then obtained all of these users' Twitter bios and self-declared locations. For users with a non-empty profile location string, we ran them through Yahoo!\ Placemaker~\footnote{\url{http://developer.yahoo.com/boss/geo/}} to map their locations to countries. We then kept 2,000 users with the appropriate country.

For each of the sampled followers we obtained the full list of their Twitter friends, i.e., users that they follow. The seed accounts pertaining to the corresponding rivalry/group were removed from these friends lists and the remaining ones were treated as a feature vector with each dimension corresponding to a Twitter account being followed. Users who followed \emph{only} seed accounts were dropped.

We used this data to construct a binary classifier for which we created train and test splits, each consisting of $\sim$1,000 users. Note that even though 2,000 users were sampled, due to limitations in the Twitter API, the actual number of users for which we could get the friends varies between 1,800-2,000. This could be due to changes in users' privacy settings, accounts getting blocked and so on.
For constructing the training vectors, we only considered users who were followed by at least two users in our training set. That is, if only one of the thousands of users in the training set followed @phdcomics, then following @phdcomics would not be used as a feature. This serves as a simple method for reducing the dimensionality as well as removing unimportant dimensions. This is analogous to the text mining scenario of removing rare tokens with a frequency 1.

%

\section{Co-Following and Binary Preferences}\label{sec:binary}
In this section we look at how much a user's choice of Twitter friends reveals about their preference among two alternatives such as @CocaCola vs.\ @Pepsi. We do this with different research questions in mind.
First, we approach things from a machine learning perspective with an evaluation of the corresponding binary classification task. Next, we do a feature analysis to see which arguably irrelevant features, such as musicians followed, provide information about a user's soft drink or political preferences. 
Finally, in Section~\ref{sec:binary:middle}, we look at whether we can identify ``the middle'' or ``undecided users'' who follow \emph{both} alternatives. Identifying this users segment is of particular relevance for political advertising.

\emph{Feature Vectors Using IDF.}
As a preprocessing step, we transformed our binary X-follows-Y vectors to an IDF-weighted alternative. To illustrate why, imagine that almost everybody follows @SuperCelebrity. Then following @SuperCelebrity is not very informative or discriminative and is given a very low IDF weight.
For each of the 18 rivalries, we compute the IDF scores of the friends of the followers of the 36 seed rivals. In total, 63,853 (N) followers of the seed accounts were used (= 36 x 2,000, minus cases with fewer than 2,000 followers and blocked/deleted/private accounts). We computed the IDF of each of their friends, obtaining one IDF-weighted vector for each of the 63,853 followers.

\begin{equation}
IDF(user_i) = log\left(\frac{N}{|followers(user_i)|}\right),
\end{equation}

where $followers(user_i)$ indicates the followers of a particular user, from the set of followers sampled for the seed rival accounts.
Each of these IDF-weighted vectors was then normalized in 2-norm and, for a given seed account, all of its followers normalized vectors are summed up. This summed vector is then re-normalized in 2-norm to give the final ``global'' summary vector for the seed account.

\subsection{Machine Learning Performance}\label{sec:binary:machinelearning}
In this section we evaluate how much information co-following provides for the task of classifying users according to their binary preference. The ground truth is the single
\footnote{We will later in Section~\ref{sec:binary:middle} look at those users individually that follow \emph{both} alternatives.} 
account that the test user actually followed. The feature dimensions corresponding to the following of the seed accounts are always removed and, later, we also remove strongly correlated features such as following @BarackObama for the @GOP vs.\ @TheDemocrats task.
Empty vectors, after removing the seed accounts (for users following only the seed accounts) are ignored.
Our main performance measure is the area-under-curve (AUC) for the Response Operating Characteristic Curve (ROC) as computed by\footnote{\url{http://mark.goadrich.com/programs/AUC/}} \cite{davisgoadrich06icml}. We also report AUC for the Precision-Recall Curve (PR) though AUC-ROC will be the default. A value of 0.5 indicates a random, unskilled prediction model.

Note that we are more interested in understanding the relative performance when, say, the most discriminative features are removed than we are in achieving the highest possible classification accuracy. 
The accuracy could always be improved further by using other algorithms (SVM, Maximum Entropy, etc.), other feature sets (textual data, interaction features, network features, etc.), or incorporating other techniques (label propagation, community detection, etc.).
Our focus is more on understanding issues such as robustness under feature removal, relative performance on sparse test vectors or opportunities for Computational Social Science arising from feature analysis.

\emph{Global vs.\ Local Approach.}
To determine whether you fall into group A or B, is it more useful to know (i) what the general, average members of A and B are like, or (ii) which of any of the two contains a small number of members just like you? The answer to this question has applications both for the design of classification algorithms and for understanding the structure of groups of followers. 
We try to answer this question by comparing two different classification strategies.
First, a ``global'' method using the single IDF-weighted summary vector described above. This method also includes information about fairly rare friends as it aggregates information from about 2,000 followers.
Second, a ``local'' approach, that uses a k-nearest neighbor classifier. It then assigns each test vector to the class with the largest number of close neighbors among the top k.
For k-NN we experimented with a range of values for k from 1 to 9 in increments of 2. There was a clear tendency for higher values of k to perform better and so we stuck to a choice of k=9. We did not experiment with larger values as the general trend of a more and more global approach performing better was our main objective, rather than identifying an optimal value of, say, k=135.

The performance of the binary classification is shown in Table~\ref{tab:performance}. The global approach always performs better than the local approach, showing that the it is worth aggregating the long tail of rare co-follower relations. 
The AUC-ROC averaged across the 18 tasks is 0.81.


\begin{table}[ht]
\centering
\begin{tabular}{p{5cm}ll}
Rivalry & Global & Local \\ \hline
@Budweiser vs.\  @MillerCoors & 0.86 (0.91) & 0.80 (0.85) \\
@FedEx  vs.\  @UPS  & 0.73 (0.73) & 0.69 (0.72) \\
@GM    vs.\   @Ford & 0.75 (0.86) & 0.69 (0.76) \\
@GOP   vs.\    @TheDemocrats & 0.91 (0.95) & 0.86 (0.93) \\
@Hertz    vs.\  @Avis & 0.92 (0.93) & 0.91 (0.92) \\
@InsideFerrari   vs.\  @lamborghini  & 0.92 (0.95) & 0.87 (0.93) \\
@jcpenney   vs.\   @Sears & 0.75 (0.82) & 0.67 (0.72) \\
@McDonalds  vs.\   @BurgerKing & 0.78 (0.79) & 0.68 (0.70) \\
@MercedesBenz  vs.\   @bmw & 0.89 (0.93) & 0.86 (0.91) \\
@Nike   vs.\  @Reebok & 0.78 (0.74) & 0.73 (0.68) \\
@NikonUSA   vs.\  @CanonUSAimaging & 0.83 (0.85) & 0.78 (0.83) \\
@pepsi   vs.\   @CocaCola & 0.69 (0.76) & 0.65 (0.73) \\
@PUMA    vs.\    @adidas & 0.77 (0.84) & 0.69 (0.73) \\
@SamsungMobile   vs.\  @TheAppleInc & 0.95 (0.96) & 0.92 (0.94) \\
@Starbucks   vs.\   @DunkinDonuts & 0.80 (0.87) & 0.72 (0.82) \\
@Target  vs.\   @Walmart & 0.78 (0.86) & 0.69 (0.79) \\
@thewanted  vs.\   @onedirection & 0.79 (0.88) & 0.76 (0.84) \\
@Visa   vs.\    @MasterCard & 0.71 (0.72) & 0.62 (0.59) \\

\end{tabular}\caption{Performance comparison for the 18 binary classification tasks (detecting preference among rivaling alternatives) in terms of AUC-ROC (AUC-PR) for both the global and local similarity-based approaches.}\label{tab:performance}
\end{table}

Note that the performance is arguably bounded by the fraction of fake accounts as such accounts might not obey any meaningful co-following behavior. According to \footnote{\url{http://fakers.statuspeople.com/Fakers/Scores}}, the typical fraction of fake followers for our 36 accounts is 20-30\% and it goes up to 50\%/39\% for @CocaCola/@pepsi, the worst-performing pair in our set, whereas it is only 5\%/12\% for @Hertz/@Avis with one of the best performances. However, for @FedEx/@UPS the classification performance is also comparatively poor, despite the low fake follower fractions of 12\%/13\%. This indicates that fake followership is not the only element influencing classification performance and that certain tasks are intrinsically harder.

%
%
%
%
%
%
%

\emph{Removing Obvious Co-following Signals.} \label{sec::binary:ranking}
Discovering that following @BarackObama on Twitter is an indication for following @TheDemocrats rather than @GOP is obvious. Similarly, following @CokeZero correlates positively with following @CocaCola. As we were more interested in studying the \emph{non-obvious} dependencies we investigated the classification performance when the most predictive features are removed. Note that this is the \emph{opposite} of what normal feature selection does.

Concretely, for each binary setting we rank features as follows. For each rivalry pair A, B, we look at the absolute differences in the feature values listed in A's and B's summary vectors. These absolute differences are then sorted in descending order. 
Features with a large difference correspond to accounts that are typically more followed by the followers of one seed account, but not the other one. 
Pseudo-code to compute the difference is shown in Algorithm~\ref{alg:pseudocode_feature_ranking}. We then rank these features by the absolute difference.

\begin{algorithm}
\caption{Pseudocode for computing difference between features}
\label{alg:pseudocode_feature_ranking}
\begin{algorithmic}
\For{$feature \in train\_A.features$} 
\If{$feature \in train\_B.features$}
\State $diff \gets train\_A[feature] - train\_B[feature]$
\Else
\State $diff \gets train\_A[feature]$
\EndIf
\EndFor
\For{$feature \in train\_B.features$} 
\If{$feature \notin train\_A.features$}
\State $diff \gets -train\_B[feature]$
\EndIf
\EndFor \\
\Return $diff$
\end{algorithmic}
\end{algorithm}

\begin{figure}[ht]
\includegraphics[width=8.5cm, clip=true, trim=10 5 25 50]{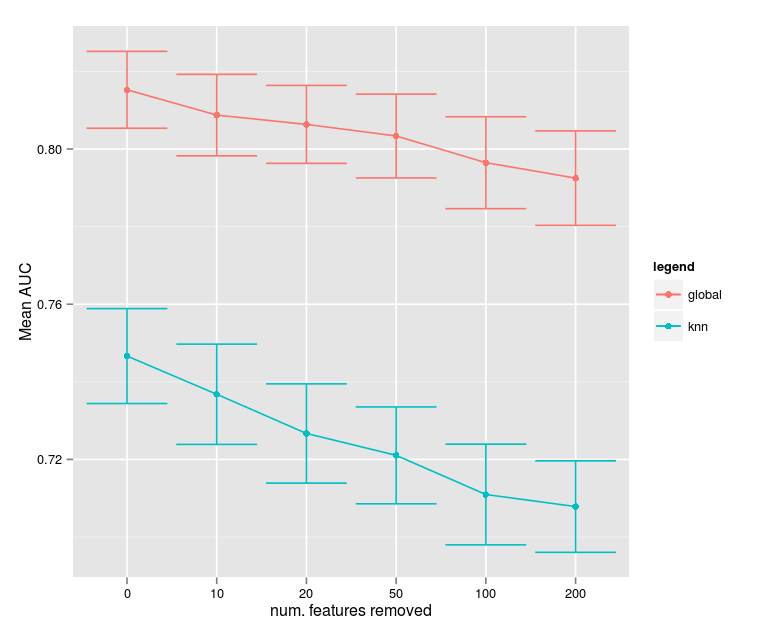}
\caption{Average AUC-ROC across the 18 binary classification tasks (detecting preference among rivaling alternatives) as more and more features are removed, but the test set size is fixed. Only users with more than 201 followers are considered. Error bars indicate the standard error across the tasks.}\label{fig:featureremoval_fixed_test}
\end{figure}

In order to check the influence of the top features, we removed the top 10, 20, 50, 100 and 200 most obvious features and compared the AUC in each case. Note that in this setting, since we remove the most influential features, the size of the test set might change (because some users might only follow these influential users). In order to compensate for this, we tried two variants, one considering only users who have more than 201 followers, so that the size of the test set is fixed and the other with a varying test set size. The results of the former case are presented in Figure~\ref{fig:featureremoval_fixed_test}, though results in both cases are comparable. The y-axis indicates AUC averaged across all the rival groups. We note a gradual decrease in the mean AUC as we remove more features, which is in line with what is expected.


For our later ``mapping'' analysis (Section~\ref{sec:mapping}), we take a similar approach to remove the top 20 features for each of the A vs.\ non-A classification problems, where A iterates over all the seed accounts and the non-A group pools all non-A seeds.




\emph{Influence of Number of People Followed.}
Different users reveal different amounts of their preferences on Twitter. Some choose to follow hundreds of users and some follow none. How much following information is needed to gauge a user's preference? Does the number of users you follow affect the performance of the classification? 
We address these questions here by bucketing the users in the test set according to the number of their Twitter friends. Apart from the seed users themselves, the 20 most discriminative (and obvious) features were ignored (see previous discussion) and only users following at least one other user were considered.
\begin{figure}[ht]
\centering
\includegraphics[width=8.5cm]{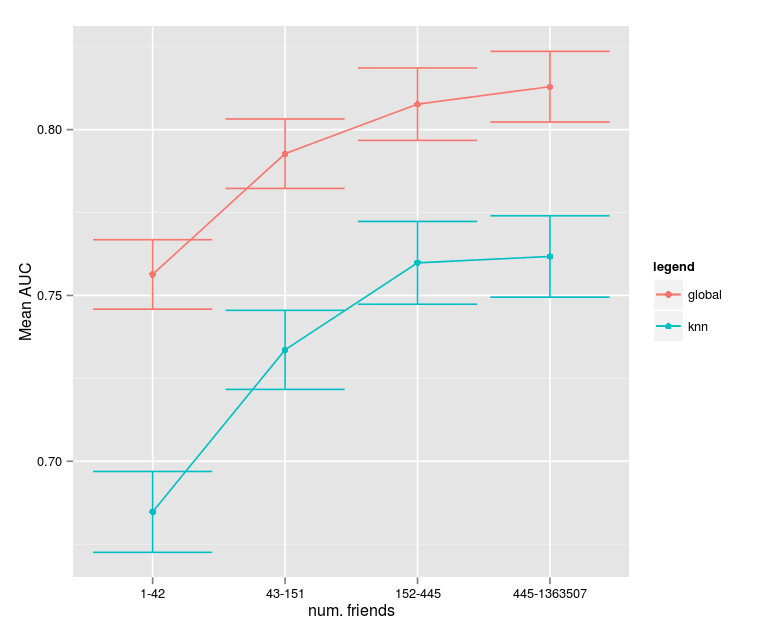}
\caption{Average AUC-ROC across the 18 binary classification tasks (detecting preference among rivaling alternatives) for different sets of users. Though the AUC increases for users with more Twitter friends, the growth slows down already when the number of users crosses 43, when the global similarity based approach is used.}\label{fig:userbuckets}
\end{figure}
%
We bucketed the test users according to the number of friends they have. In order to make qualitative conclusions over the general population, we computed these buckets by pooling the followers of all seed users. We then found the $25^{th}$, $50^{th}$, $75^{th}$ and $100^{th}$ percentiles for the number of friends these users have and bucketed the users for each group accordingly. The buckets were $bucket1 \in [1,43)$, $bucket2 \in [43,152)$, $bucket3 \in [152,446)$, and $bucket4 \in [446,1,363,508]$. Figure~\ref{fig:userbuckets} plots the average AUC across the four user buckets. As expected, the more followers you have, the easier it is to know your relative preference. An interesting observation is that for the ``global'' approach, the curve almost subsides after the first bucket, indicating that the value added after 43 followers is not noticeable. It also indicates that $\sim$100 users is sufficient to predict the class of a user with $>$ 0.8 AUC-ROC.

\subsection{Feature Analysis}\label{sec:binary:features}

\emph{Correlation with Lifestyle.} 
Apart from using it for user classification and target advertising, co-following patterns are also of interest in their own right and can serve to answer questions in Computational Social Science. For instance, there is academic work that looks at ``lifestyle politics'' such as the correlation between political leaning and television preferences or the stereotype that liberals like lattes \cite{zogby08,bernsteinetal06forum,postaetal13submitted}. Our approach contributes to this by offering a language-agnostic tool to use online data to quantify such effects at scale.
In this section, we use the feature ranking described previously to generate the top, discriminative features (such as following @BarackObama to predict a @GOP or @TheDemocrats preference).

As an example, we look at the rivalries between @GOP vs.\ @TheDemocrats, and @Pepsi vs.\ @Cocacola. 
For both cases we look at the top discriminative co-following features from the \url{http://WeFollow.com} categories Music, Sports and News~\footnote{A small number of mislabeled entries were removed. For example, @AgainAmerica was incorrectly listed in the Music category.}.
Table~\ref{tab:differences} shows some examples of the insights we can get from co-following patterns. The lifestyle correlations for the political rivalry @GOP vs.\ @TheDemocrats can be inspected to make intuitive sense with, e.g., @nytimes being more popular among @TheDemocrats followers.\footnote{The New York Times is generally perceived to have a liberal bias, see \url{http://en.wikipedia.org/wiki/The_New_York_Times\#Political_persuasion_overall}.} 
For @Pepsi vs.\ @CocaCola many observations can be explained by the fact that @Pepsi targets the younger ``New Generation''. 
 In passing we also mention that following @Starbucks is indeed an indicator for co-following @TheDemocrats rather than @GOP.

\begin{table}[ht]
\centering
\begin{tabular}{ccc}
& @GOP & @TheDemocrats \\ \hline
Music 1 & @kennychesney (64) & @ladygaga (24) \\
Music 2 & @jakeowen (122) & @aliciakeys (57) \\
Music 3 & @taylorswift13 (139) & @SnoopDogg (62) \\ \hline
Sports 1 & @espn (125) & @rolandsmartin (66) \\
Sports 2 & @runnersworld (143) & @bubbawatson (79) \\
Sports 3 & @AdamSchefter (178) & @NBA (188) \\ \hline
News 1 & @WSJ (12) & @nytimes (11) \\
News 2 & @HumanEvents (77) & @cnnbrk (21) \\
News 3 & @toddstarnes (107) & @NYTimeskrugman (23) \\ \hline \hline
& @Pepsi & @Cocacola \\ \hline
Music 1 & @SnoopDogg (1) & @maroon5 (8) \\
Music 2 & @Nickiminaj (2) & @davidguetta (19) \\
Music 3 & @Drake (5) & @Pitbull (28) \\ \hline
Sports 1 & @shaq (3) & @SInow (99) \\
Sports 2 & @ochocinco (20) & @kaka (103) \\
Sports 3 & @DwightHoward (42) & @chicagobulls (206) \\ \hline
News 1 & @Rapup (120) & @cnnbrk (55) \\
News 2 & @Life (133) & @WSJ (81) \\
News 3 & @MTVnews (339) & @TheOnion (183) \\ \hline
\end{tabular}\caption{List of differentiating co-following features from different WeFollow classes, for @GOP vs.\ @TheDemocrats, and @Pepsi vs.\ @Cocacola. 
 The numbers in parentheses indicate the absolute position of this feature in our ranking irrespective of the topic (Music, Sports or News).}\label{tab:differences}
\end{table}

\emph{Celebrity Endorsements.}
An interesting side note of examining the top features is the detection of celebrity endorsements. 
By observing these features, we found out that celebrity endorsements go hand in hand with who people follow. 
Some examples include Derrick Rose (@drose) for Adidas; Kevin Durant (@KDTrey5) for Nike; SnoopDogg (@SnoopDogg), Nicki Minaj (@NickiMinaj) and Drake (@drake) for Pepsi, and Maroon 5 (@maroon5) and David Guetta (@davidguetta) for Cocacola, and not so famous relations like Dan Bailey (@Dan\_Bailey9) and Rich Froning (@richfroning) to Reebok. (They are both stars of the show CrossFit Games, hosted by Reebok.) 
This observation has interesting applications in marketing campaigns and recommendation systems and deserves more analysis in the future.


\subsection{Analyzing the ``Center''}\label{sec:binary:middle}
For certain marketing campaigns, in particular in the political domain, identifying strong supporters or sworn enemies is less useful as identifying users in the likely \emph{center}. In this study, we define the center as users that follow \emph{both} of two alternative and rivaling accounts. We investigate whether this center is then indeed ``in the middle'' and whether such users can be told apart from the more polarized users.

\begin{figure}[ht]
\centering
\includegraphics[width=6cm]{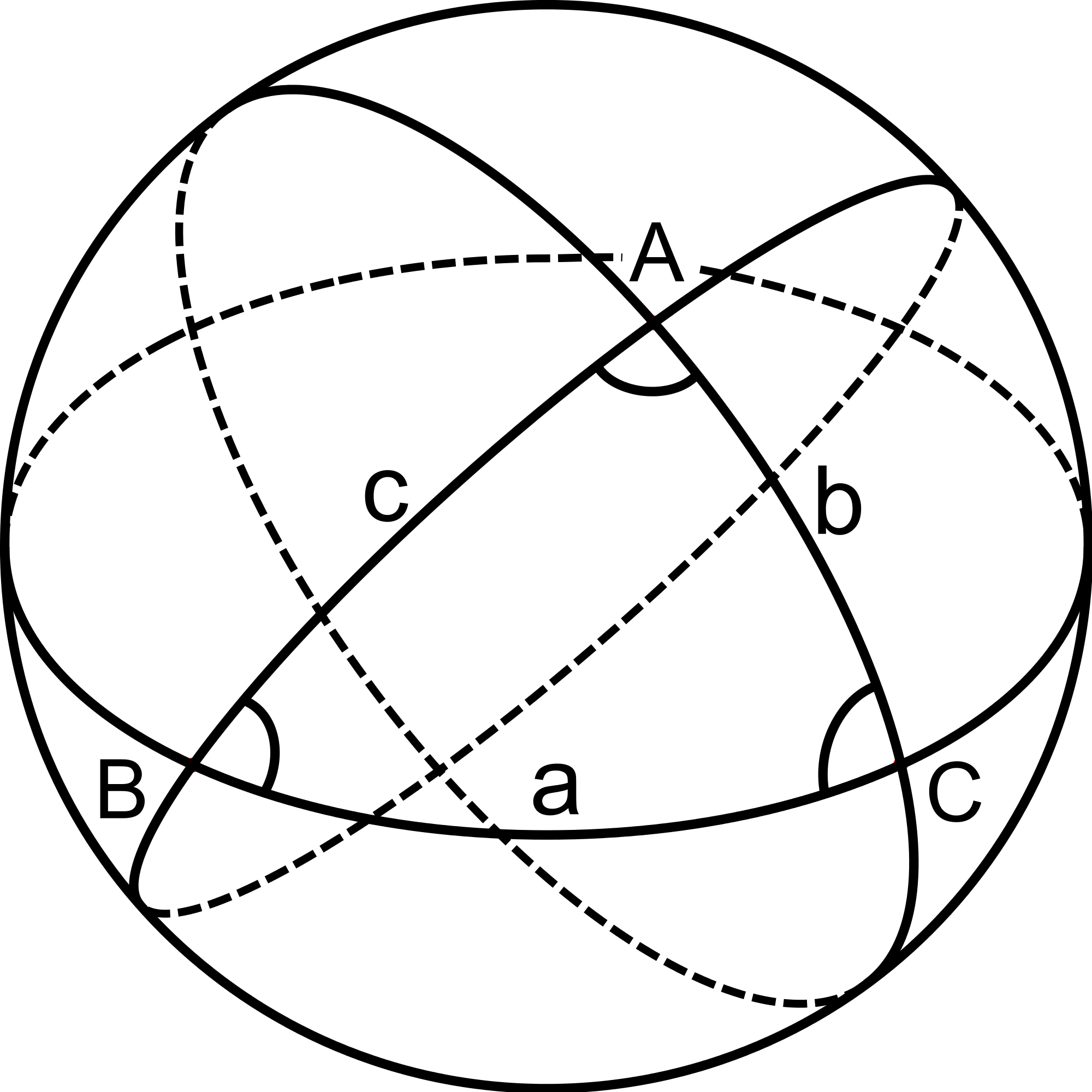}
\caption{Illustration of basic spherical trigonometry that is used to quantify the degree of centered-ness of the dual-followership crowd (centroid $C$) with respect to the two alternatives' centroids ($A$ and $B$).}\label{fig:spherical}
\end{figure}

\emph{Is the Center in the Middle?}
One definition to lie ``in the middle'' is that the distance to either end of the spectrum should be smaller than the distance from one end of the spectrum to the other. If we think of the triangle formed by the two alternatives $A$ and $B$ and the supposed center $C$ then we require that both of the lengths $|AC|$ and $|CB|$ are smaller than $|AC|$. In our case, we are looking at 2-normalized vectors lying on the unit sphere and so we require that both $b=\angle(A,C) < c=\angle(A,B)$  and  $a=\angle(C,B) < c=\angle(A,B)$. Figure~\ref{fig:spherical} shows an illustration where $A$ and $B$ refer to the unit vectors of the centroids of the two corresponding rivals, with $C$ referring to the centroid of the dual-followership. The results of this test of the angles which we refer to as $\Delta$ is presented in Table~\ref{tab:middle}. For 12 of the 18 cases this condition is satisfied, but for six cases it is violated. This gives evidence that in general users following two similar but rivaling alternatives are indeed \emph{between} the two ends, but due to the small number of rivalries considered the result is not statistically significant. In future work we plan to look more at when this fails to hold and when these users are not between but rather \emph{different}.
To quantify how much in the middle the dual-followership is, we also look at the angle of the spherical triangle formed at $C$. See Figure~\ref{fig:spherical} for an illustration. The larger the angle at $C$ the closer to the middle $C$ is. This angle is listed in Table~\ref{tab:middle} and it indicates that for @pepsi vs.\ @CocaCola $C$ is furthest away from the middle whereas for @Hertz vs.\ @Avis it is closest. Note that in no case the angle formed at $C$ exceeds $90^\circ$ indicating that $C$ is both partly a ``bridge'' in the middle but still largely a class by itself. We will explore this in detail in the next paragraph.
%
%
%
%
%
\begin{table}[ht]
\centering
\begin{tabular}{ccc}
Rivalry & $\Delta$ & angle at C$^\circ$ \\ \hline
@Budweiser vs.\  @MillerCoors & Yes & 72 \\
@FedEx  vs.\  @UPS  & Yes & 65 \\
@GM    vs.\   @Ford & Yes & 66 \\
@GOP   vs.\    @TheDemocrats & Yes & 64 \\
@Hertz    vs.\  @Avis & Yes & 76 \\
@InsideFerrari   vs.\  @lamborghini & Yes & 65 \\
@jcpenney   vs.\   @Sears & Yes & 68 \\
@McDonalds  vs.\   @BurgerKing & No & 64 \\
@MercedesBenz  vs.\   @bmw & Yes & 68 \\
@Nike   vs.\  @Reebok & No & 64 \\
@NikonUSA   vs.\  @CanonUSAimaging & No & 68 \\
@pepsi   vs.\   @CocaCola & No & 58 \\
@PUMA    vs.\    @adidas & Yes & 64 \\
@SamsungMobile   vs.\  @TheAppleInc & Yes & 66 \\
@Starbucks   vs.\   @DunkinDonuts & Yes & 66 \\
@Target  vs.\   @Walmart & Yes & 66 \\
@thewanted  vs.\   @onedirection & No & 66 \\
@Visa   vs.\    @MasterCard & No & 62 \\
\end{tabular}\caption{Results of whether users following \emph{both} alternatives can be considered to lie ``in the middle''. $\Delta$ refers to the necessary condition of the angles, see text.}\label{tab:middle}
\end{table}

\emph{Can We Tell the Middle From the Poles?}
In the previous paragraph we saw both evidence for the dual-followership crowd \emph{lying between} the two ends (as $\Delta$ was satisfied in 12 out of 18 cases) but also for being a \emph{qualitatively different} class of their own (as the angle at $C$ never approached $90^\circ$. In this paragraph we look at whether the dual-followership can be told apart from the two poles and which features are of relevance in this process.
For the classification, we compute the three centroids for the two rivals and dual-followership following the same procedure as before. Then for the $\sim$2,000 test instances, roughly 500 from each pole and 1,000 from the dual group, we looked at with which centroid they shared the highest similarity. For Table~\ref{tab:middle_performance} we look only at the Both-or-Not performance where the two rivals would be treated as one class.
\begin{table}[ht]
\centering
\begin{tabular}{ccc}
Rivalry & Global & Local \\ \hline
@Budweiser vs.\  @MillerCoors & 0.88 (0.91) & 0.87 (0.85) \\
@FedEx  vs.\  @UPS & 0.79 (0.74) & 0.81 (0.81) \\
@GM    vs.\   @Ford & 0.73 (0.79) & 0.77 (0.79) \\
@GOP   vs.\    @TheDemocrats & 0.78 (0.72) & 0.89 (0.86) \\
@Hertz    vs.\  @Avis & 0.83 (0.91) & 0.77 (0.84) \\
@InsideFerrari   vs.\  @lamborghini & 0.79 (0.78) & 0.84 (0.82) \\
@jcpenney   vs.\   @Sears & 0.78 (0.76) & 0.81 (0.81) \\
@McDonalds  vs.\   @BurgerKing & 0.77 (0.74) & 0.85 (0.85) \\
@MercedesBenz  vs.\   @bmw & 0.73 (0.73) & 0.80 (0.78) \\
@Nike   vs.\  @Reebok  & 0.77 (0.69) & 0.85 (0.82) \\
@NikonUSA   vs.\  @CanonUSAimaging & 0.79 (0.75) & 0.84 (0.83) \\
@pepsi   vs.\   @CocaCola & 0.84 (0.79) & 0.90 (0.87) \\
@PUMA    vs.\    @adidas & 0.91 (0.91) & 0.91 (0.88) \\
@SamsungMobile   vs.\  @TheAppleInc & 0.78 (0.74) & 0.86 (0.84) \\
@Starbucks   vs.\   @DunkinDonuts & 0.81 (0.84) & 0.82 (0.80) \\
@Target  vs.\   @Walmart & 0.76 (0.78) & 0.78 (0.77) \\
@thewanted  vs.\   @onedirection & 0.82 (0.85) & 0.76 (0.73) \\
@Visa   vs.\    @MasterCard & 0.77 (0.81) & 0.77 (0.81) \\
\end{tabular}\caption{Performance comparison for the 18 binary classification tasks (detecting the ``center'' from the two poles) in terms of AUC-ROC (AUC-PR) for both the global and local similarity-based approaches.}\label{tab:middle_performance}
\end{table}

Looking more closely at the relative closeness of the three classes, we also looked at the three-class classification problem and the corresponding confusion matrices. Concretely, we looked at the number of rival pairs A-B for which both \#(true A, closest to C) > \#(true A, closest to B) and \#(true B, closest to C) > \#(true B, closest to A), where C corresponds to the dual-followership. Intuitively, if both of these conditions are satisfied for a pair then C is the most confusing class, indicating it is between the others. Surprisingly, for only three out of the 18 cases this was the case. The story of when dual-followership indicates a ``middle ground'' therefore seems to be complex.
\section{Mapping the Twittersphere via Co-Following}\label{sec:mapping}
In this section we look at whether a co-following based similarity can be used to map the relative positions of players from domains as diverse as politics and music. 
Though our maps can be seen as ``community detection'', the approach and interpretation is very different. Whereas traditional community detection algorithms use direct social links and, e.g., would try to find clusters with unusually high triadic closure \cite{newman06pnas}, our approach relies on more indirect and high-order links. As an example, imagine two football clubs that are fierce rivals and who would definitely not follow each other. Fans and Twitter followers of either club might also not follow the other one. However, their fans might jointly follow many other accounts related to sports news. Due to this co-following of the clubs' followers we would consider the clubs as similar in terms of their audiences' interests.
This approach also opens up opportunities for cross-marketing and cross-selling: if two Twitter accounts from different domains share a similar followership then they might consider cross-posting or otherwise combining their forces. Note again that they do not have to share even a single follower to be considered similar as we look at second-order following relations, namely, the friends of their followers.

Technically, we did the following: For each of the $\sim$2,000 followers of an account, we constructed the IDF vector for the users they follow.\footnote{The total number of followers (N) was 38,358 (Musicians), and 7,783 (Germany), and 9,966 (France).}.
 We then computed the pair-wise cosine similarity between these feature vectors. Since we need distances, we used (1 - cosine similarity) as the measure of distance. We then used the classical, Metric Multi-Dimensional Scaling (MDS)~\cite{liwen06jasist} on this data with the $cmdscale()$ function in $Matlab$.

Note that MDS is a \emph{lossy} embedding and that even though two points appear close in the 2-dimensional plane, they might be far apart in the original high dimensional space. Therefore, all conclusions and observations we derived from such mappings in the following have also been validated using the high dimensional similarity information. 





\subsection{Political Parties in Germany and France} \label{sec:mapping:france_germany}
Both for France and for Germany we obtained a list of parties/alliances that were present in the corresponding national parliaments with a non-trivial number of seats.
We were interested to see if the grassroot-derived party similarities matched the ideological realities.
For France our list included:
Europe \'{E}cologie Les Verts/Europe Ecology The Greens (@EELV, center-left, 104,997 followers), Front de Gauche/Left Front (@FDG, left, 17,481 followers), Union des D\'{e}mocrates et Ind\'{e}pendants/Union of Democrats and Independents (@UDI\_off, center, 7,825 followers), 
Union pour un Mouvement Populaire/Union for a Popular Movement (@ump, center-right, 75,292 followers), Parti socialiste/Socialist Party (@partisocialiste, center-left, 67,693 followers), and Front national/National Front (@FN\_officiel, far right, 25,429 followers).

For Germany we used:
Die Linke/The Left (@dieLinke, left, 22,746 followers), Freie Demokratische Partei/Free Democratic Party (@fdp\_de, center-right, 17,006 followers), Die Gr\"{u}nen/The Greens (@Die\_Gruenen, center-left, 72,920 followers), Christlich Demokratische Union \& Christlich Soziale Union/Christian Democratic Union \& Christian Social Union (@cducsubt, center-right, 19,998 followers), Sozialdemokratische Partei Deutschlands/Social Democratic Party of Germany (@spdde, center-left, 45,407 followers).

\begin{figure}
\centering
\includegraphics[width=.23\textwidth, clip=true, trim=10 10 25 10]{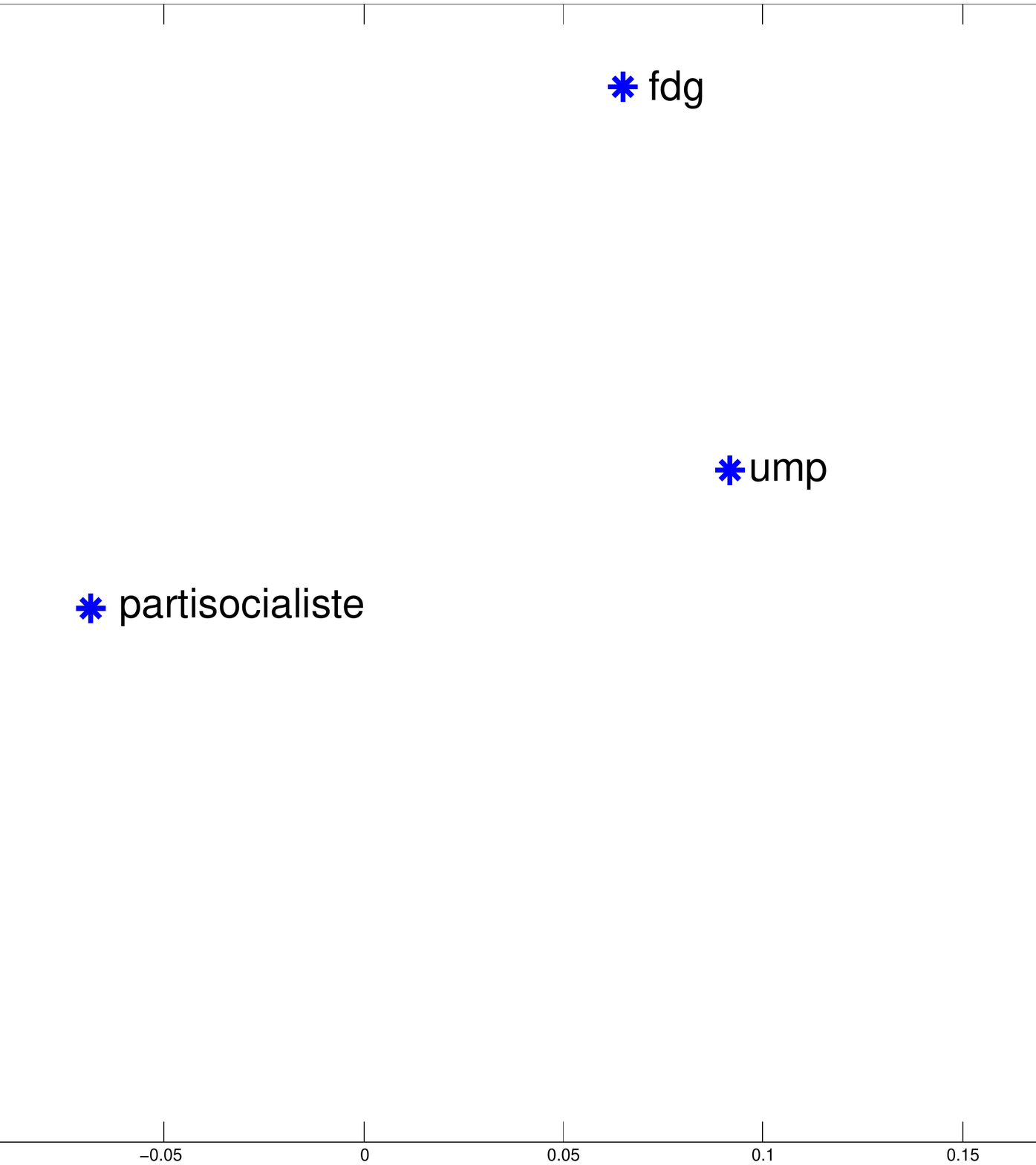}
\includegraphics[width=.23\textwidth, clip=true, trim=50 50 25 10]{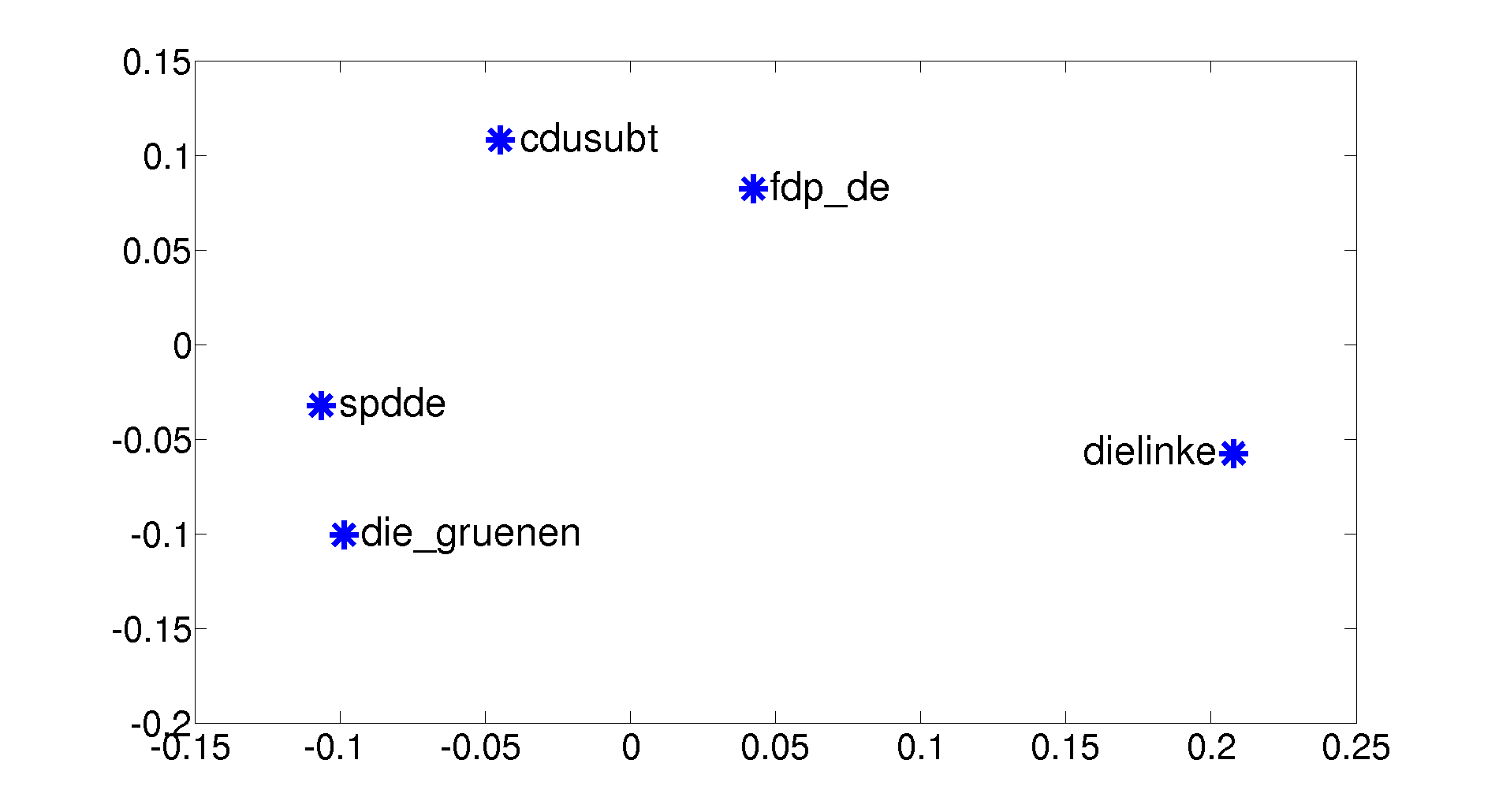}
\caption{A 2D MDS similarity map of French and German political parties. Similarity measures are derived from their followers' aggregated friends.}
\label{fig:france_germany}
\end{figure}



For both countries the results make intuitive sense. For France (Figure~\ref{fig:france_germany}, left) the Front de Gauche (@fdg) and the Front national (@fn\_officiel) are the two parties furthest away from the center in terms of political spectrum, whereas the other center-left and center-right parties are close to each other. Similarly, for Germany (Figure~\ref{fig:france_germany}, right) 
, the isolated outlier is Die Linke (@dielinke), a party which is frequently boycotted/ignored by other parties in terms of coalition considerations, and the remaining parties are close to their respective frequent coalition partners, namely, Die Gr\"{u}nen (@die\_gruenen) \& Sozialdemokratische Partei Deutschland (@spdde); Freie Demokratische Partei (@fdp\_de) \& Christlich Demokratische/Soziale Union (@cducsubt). 
Note that it is not obvious that the similarity of parties' followers should agree with the parties' orientation as the majority of a follower's Twitter friends will be apolitical.
Also see Section~\ref{sec:binary:features} for examples of ``lifestyle politics'' in the case of the US.

\subsection{Popular Musicians}\label{sec:mapping:musicians}
To see if our approach generalizes to other domains, we decided to map popular musicians on Twitter.
To this end we obtained a list of the top 22 musicians (in terms of prominence score) from~\url{http://wefollow.com/interest/music}.\footnote{We removed 4 accounts from the initial list that corresponded to media/producers rather than musicians or bands.}

\begin{figure}[ht]
\centering
\includegraphics[width=8.5cm, clip=true, trim=40 50 25 10]{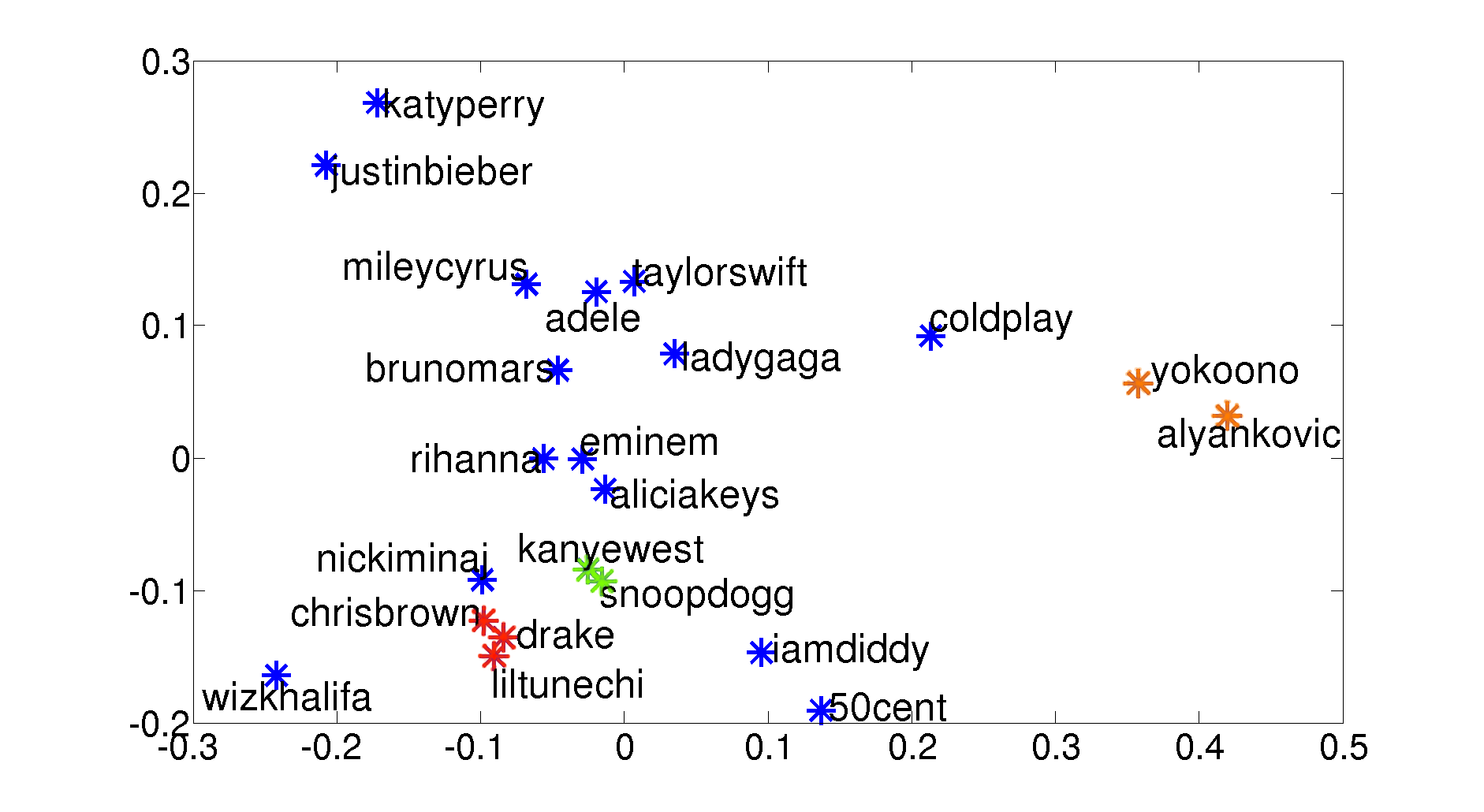}
\caption{A 2D MDS similarity map of popular musicians. Similarity measures are derived from their followers' aggregated friends.}\label{fig:musicians}
\end{figure}

Figure~\ref{fig:musicians} shows the map that was obtained in the same way as as the previous case. Most of the observed structure corresponds to musical genres. For example, Lil Wayne (@liltunechi), Chris Brown (@chrisbrown) and Drake (@drake) are rappers and are co-mapped together in the map, marked in red. Similar is the case of Snoop Dogg (@snoopdogg) and Kanye West (@kanyewest), marked in green, both of which are hip hop artists. However, there are also surprising things that emerge such as the relative closeness of ``Weird Al'' Yankovic (@alyankovic), famous for musical parody, and Yoko Ono (@yokoono), both marked in orange. Though very different musical genres, both arguably appeal to an older, more educated audience. This already hints at applications of such analysis for the identification of cross-selling opportunities.

\subsection{Combination of All Rivals}\label{sec:mapping:all_rivals}
\begin{figure}[ht]
\centering
\includegraphics[width=8.5cm, clip=true, trim=30 50 25 10]{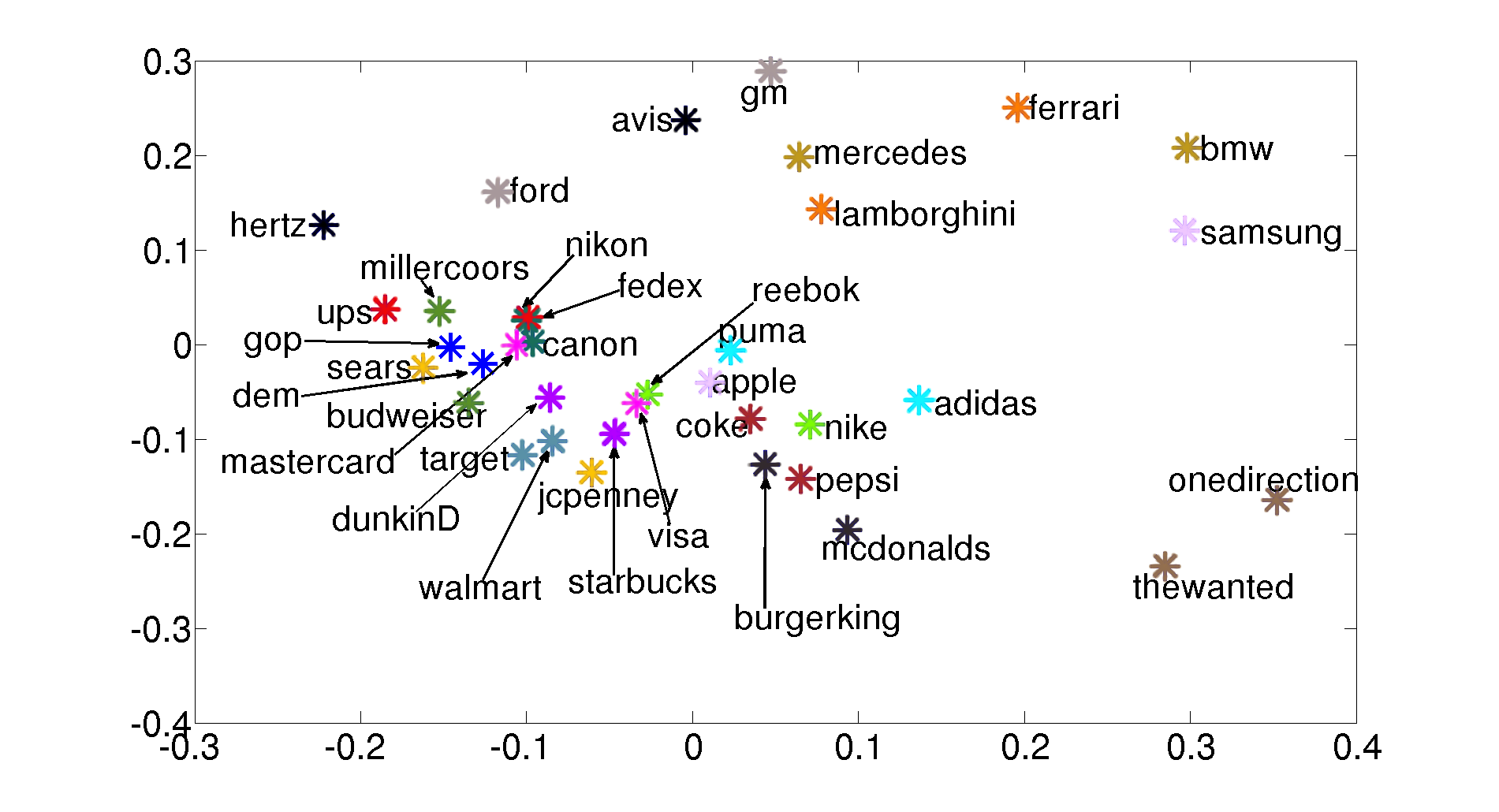}
\caption{A 2D MDS similarity map of all the 18x2 rivals. Similarity measures are derived from their followers' aggregated friends. Rival pairs are represented by stars of the same color. Some labels have been shortened due to space constraints. dunkinD represents DunkinDonuts and dem., Democrats.}\label{fig:all_rivals}
\end{figure}

To show the full generalizability of this mapping approach also \emph{across} domains we combined all the 36 Twitter accounts from the 18 rivalries and mapped them in a common space in Figure~\ref{fig:all_rivals}. 
As one would expect, many rivals such as @Target vs.\ @jcpenney and @thewanted vs.\ @onedirection are comparatively close as their followers share similar interests. However, the relative distances across rivalries also makes sense. 
For example, the beer brand @MillerCoors is closer to @GOP than to @TheDemocrats and the opposite holds for @Budweiser. This makes sense as it has been observed before that ``Republicans are also big fans of Miller Lite and Coors Light, but Democrats drink more Budweiser'' \cite{doyle12boston,shannonfeltus12national}, though, some studies show that the opposite is true~\cite{time2013beer}. Sometimes, studies such as this one are inconclusive and show conflicting results based on the demographics studied (such as voters vs. just politically leaning but not necessarily voting), sampling methods used, etc.  Similarly, @GM and @Avis are very close in the low-dimensional embedding. Again, this makes sense as ``[s]ince the late 1970s, Avis has featured mainly General Motors (GM) vehicles''\footnote{\url{http://en.wikipedia.org/wiki/Avis_Rent_a_Car_System}, accessed on Mar 20, 2014.}. Also noteworthy is the closeness between @PUMA and @TheAppleInc. Though there does not appear to be any formal alliance, both brands try to create a similar image of themselves. Puma targets the ``sports lifestyle'' trend with persona attributes such as metropolitan and international \cite{kahute06} which, arguable also apply to Apple.

We believe that such a mapping is useful to quickly generate hypotheses for lifestyle politics and similar research areas that can then be investigated in depth. It is important to note that even if some of these findings were not to hold ``offline'' in all cases, these Twitter-only findings are still useful for online advertising as they definitely provide a signal.

\section{When Being Mainstream is (Not) Best}\label{sec:mainstream}
Does being ``mainstream'' always equal being more popular? We try to answer this question by looking at whether Twitter accounts with the more \emph{normal} followers typically also have \emph{more} followers than their direct rivals. To establish what ``normal'' means we combine all the $\sim$63,000 followers of the 18 rivals from the US that we used in our previous analysis. All following information for any of the seed accounts is removed and the remaining non-empty following feature vectors are normalized in 2-norm, and then summed and re-normalized. This ``average centroid'' is used as a reference for the follower preferences of a normal user.
For each of the 18 pairs we then looked whether the rival whose followers were closest to this average also had more followers from the US. Here we estimated the number of followers from the US by first computing the fraction of users containing a profile location string as ``US'' out of a randomly sampled set of 10k users (see Section~\ref{sec:data}).
This fraction was then applied to the total number of followers to obtain an estimate. Note that though the total number of followers from the US is likely larger, we followed this procedure to have an estimate about the user population from which the training/test users were drawn as for them the same procedure was applied.
\begin{table*}[ht]
\centering
\begin{tabular}{cccc}
 &           & \multicolumn{2}{c}{Similarity} \\
Rival 1 vs.\ Rival 2 &  \# fwers & to aver.\ centroid & within cluster \\ \hline
@Budweiser vs.\  @MillerCoors &  9,614 / 2,524 & 0.81 / 0.76 & 0.73 / 0.64 \\    
@FedEx  vs.\  @UPS  & 27,681 / 21,585 & 0.78 / 0.78 & 0.76 / 0.71 \\   
@GM    vs.\   @Ford & 39,130 / 71,784 & 0.83 / 0.80 & 0.70 / 0.65 \\   
@GOP   vs.\    @TheDemocrats  & 99,420 / 100,008 & 0.80 / 0.82 & 0.77 / 0.81 \\  
@Hertz    vs.\  @Avis  & 10,523 / 2,214 & 0.84 / 0.83 & 0.68 / 0.51 \\  
@InsideFerrari   vs.\  @lamborghini  & 19,304 / 19,422 & 0.81 / 0.85 & 0.63 / 0.69 \\  
@jcpenney   vs.\   @Sears & 78227 / 23133 & 0.85 / 0.83 & 0.78 / 0.77 \\
@McDonalds  vs.\   @BurgerKing & 363,184 / 61,617 & 0.87 / 0.84 & 0.80 / 0.85 \\  
@MercedesBenz  vs.\   @bmw & 14,630 / 2,874 & 0.78 / 0.80 & 0.73 / 0.58 \\  
@Nike   vs.\  @Reebok  & 276,747 / 34,259 & 0.81 / 0.80 & 0.76 / 0.79 \\  
@NikonUSA   vs.\  @CanonUSAimaging  & 26,474 / 5,286 & 0.76 / 0.74 & 0.66 / 0.61 \\ 
@pepsi   vs.\   @CocaCola  & 240,481 / 127,723 & 0.83 / 0.78 & 0.81 / 0.75 \\  
@PUMA    vs.\    @adidas  & 18,860 / 24,861 & 0.79 / 0.81 & 0.69 / 0.73 \\ 
@SamsungMobile   vs.\  @TheAppleInc & 95,095 / 7,098 & 0.81 / 0.80 & 0.65 / 0.59 \\  
@Starbucks   vs.\   @DunkinDonuts  & 922,533 / 142,709 & 0.87 / 0.83 & 0.66 / 0.79 \\  
@Target  vs.\   @Walmart  & 265,424 / 125,778 & 0.81 / 0.78 & 0.78 / 0.75 \\  
@thewanted  vs.\   @onedirection  & 111,166 / 693,313 & 0.81 / 0.80 & 0.88 / 0.75 \\  
@Visa   vs.\    @MasterCard & 42,723 / 18,780 & 0.86 / 0.84 & 0.81 / 0.76 \\
\end{tabular}\caption{Results for the similarity analysis of the rivals' centroids to the average centroid. The number of followers here is an estimate of the number of followers from US.}\label{tab:mainstream}
\end{table*}
Only for three out of 18 cases the account whose followers were further away from the global average had more US followers than their rival. This gives statistically significant (p=.05) evidence that, generally, mainstream means more followers.
Related to the idea whether being mainstream pays off, we also look at whether having a homogeneous crowd of followers pays off. We quantify the degree of homogeneity by the average cosine similarity of two random followers of a particular seed account. This within-cluster similarity is shown in the last column of Table~\ref{tab:mainstream}. For 14 out of 18 case the rival with the more homogeneous followership also had a bigger followership. We also note that the similarity scores between cluster centroids and the global average centroid are higher than the within-cluster similarity scores. This again indicates that a ``global'' approach works better for this data, as we saw in Section~\ref{sec:binary:machinelearning}.

\section{Conclusions}\label{sec:conclusions}
We presented an in-depth study of co-following behavior on Twitter which contributes to (i) a better understanding of language-agnostic user classification on Twitter, (ii) eliciting opportunities for Computational Social Science, and (iii) improving online marketing by identifying cross-selling opportunities.
Concretely, we used the similarity of followers' friends to predict a users' preferences and to group main Twitter users according to their audiences' similarities. 
We showed that such language-agnostic co-following information provides strong signals for diverse classification tasks and that these signals persist even when (i) the most discriminative features are removed and (ii) only relatively ``sparse'' users with fewer than 152 but more than 43 Twitter friends are considered.
Rather than solely focusing on the classification task, we presented applications of our methodology to the area of Computational Social Science and confirmed stereotypes such as that the LBGT activist @ladygaga is more popular among @TheDemocrats followers than among @GOP followers.
In the domain of marketing we gave evidence that celebrity endorsement is reflected in co-following and we demonstrated how our methodology can be used to reveal the audience similarities between Apple and Puma and, less obviously, between Nike and Coca-Cola.  Concerning a user's popularity we found a statistically significant connection between having a more ``average'' followership and having more followers than direct rivals.
To the best of our knowledge, this is the first systematic study that shows how co-following on Twitter can be used for a variety of applications.

Our main focus in this paper was to introduce the concept of (2nd order) co-following and examine how it works for a wide range of settings, rather than the algorithm itself. In future, we would like to focus more on the algorithmic perspective and extend our work by looking deeper into aspects such as comparing how co-following fares in comparison to, e.g., methods which use the tweet content and user profile or compare the plots generated by MDS with other community detection algorithms.




\bibliographystyle{abbrv}
\bibliography{co-following}  

\end{document}